\documentclass[prd,twocolumn,showpacs,superscriptaddress,nofootinbib,floatfix,showkeys,10pt]{revtex4-2}
\usepackage{graphicx}
\usepackage{amsmath}
\usepackage{bm}
\usepackage{yhmath}
\usepackage{mathtools}
\usepackage{wasysym}
\usepackage[colorlinks,citecolor=violet,urlcolor=violet,linkcolor=blue]{hyperref}
\usepackage{color}
\usepackage{cases}
\usepackage{subfigure}
\usepackage{times}
\usepackage{dcolumn,booktabs,bm}
\usepackage{slashed}
\usepackage{amsfonts,amssymb,stmaryrd,latexsym,amsmath}
\usepackage{textcomp}
\usepackage{multirow}
\usepackage{cancel}
\usepackage{array}
\usepackage{orcidlink}
\usepackage{physics}
\usepackage{pstricks}
\usepackage{color}
\usepackage{amssymb}
\usepackage{epsfig}
\usepackage{mathtools}
\RequirePackage{graphicx}
\RequirePackage{mathptmx}

\begin{document}

\title{Unpolarized valence GPDs and form factors of pion in the modified chiral quark model}
	
	\author{H. Nematollahi\,\orcidlink{0000-0003-0162-3085}}\email{hnematollahi@uk.ac.ir}    
	\affiliation{Faculty of Physics, Shahid Bahonar University of Kerman, Kerman 76169133, Iran }

	\author{K. Azizi\,\orcidlink{0000-0003-3741-2167}}\email{kazem.azizi@ut.ac.ir}
	\affiliation{Department of Physics, University of Tehran, North Karegar Avenue, Tehran 14395-547, Iran}
		\affiliation{Department of Physics, Do\u gu\c s University, Dudullu-\"Umraniye, 34775 Istanbul, 
	             T\"urkiye}

\begin{abstract}

We calculate the valence generalized parton distribution functions (GPDs) of pion at zero skewness applying a theoretical approach in which the valence GPDs are related to valence quark distribution functions, directly. To this end, we use the results of modified chiral quark model ($\chi QM$) for the valence quark distributions of pion obtained in our previous work. We also determine the electromagnetic and gravitational form factors of pion and compare the results of our theoretical model for valence GPDs and form factors of pion with the results of some other models and available experimental data.

\end{abstract}
\maketitle
	
	\thispagestyle{empty}

\section{\label{sec-intro}Introduction}
Generalized parton distribution functions \cite{GPD1,GPD2,GPD3,GPD4,GPD5}, which were introduced more than two decades ago, provide essential information about the internal structure and three-dimensional (3D) description of hadrons. Unlike the standard parton distribution functions (PDFs) of hadrons that are the functions of the Bjorken longitudinal variable $x$, GPDs are dependent on $x$, the skewness parameter $\xi$ and momentum transfer squared $t$. Due to this higher-dimension nature of GPDs, our knowledge about them is limited in comparison with PDFs. From experimental point of view, the GPDs are accessible via some exclusive processes such as deeply virtual Compton scattering (DVCS) and deeply virtual meson production (DVMP).

Since pion as the lightest meson has an important role in the description of the structure of nucleons and nuclei and also is one of the Goldstone (GS) bosons responsible for the dynamical chiral symmetry breaking, now we focus on the GPDs of this meson.

In theory, there are some models for calculating GPDs of pion such as the Nambu-Jona-Lasinio (NJL) model \cite{NJL1,NJL2,NJL3}, lattice QCD methods \cite{LQCD1,LQCD2} and other theoretical analyses that give important insights considering the GPDs of pion, e.g., Refs. \cite{PiGPD1,PiGPD2,PiGPD3,PiGPD4,PiGPD5,PiGPD6,PiGPD7,PiGPD8,PiGPD9,PiGPD10,PiGPD11,PiGPD12,PiGPD13,PiGPD14}. Furthermore, some dynamical models of GPDs exist in which the hadron states are approximated as the expansions in the Fock space using light front wave functions (LFWFs) \cite{LFWFs1,LFWFs2,LFWFs3,LFWFs4,LFWFs5}.

On the other hand, a remarkable property of GPDs is that the form factors (FFs) are related to them via sum rules \cite{Bakulev2000,Goeke2001,Diehl2003,Belitsky2005,Boffi2007}. In fact, different FFs of hadron are obtained through the integration of different momenta of GPDs over $x$ \cite{GPD1,GPD3,GPD4}. The standard electromagnetic form factor (EMFF) and gravitational form factor (GFF) of hadrons are achieved from the lowest moments of their GPDs. Since GFF is related to some fundamental quantities such as spin and mass, its investigation has been of great interest in particle physics. The EMFF and GFF of pion have been calculated applying NJL model \cite{NJL3} and the chiral quark model \cite{PiChi1,PiChi2}. These form factors have also been studied in Refs. \cite{NJL3,PiGPD7,PiGPD8,PiGPD11,PiGPD12,PiGPD13,PiChi1,PiFF1,PiFF2,PiFF3,PiFF4,Raya2024}. It should be noted that the GPDs and FFs of nucleon have been studied in Refs. \cite{GA22,GA231,GA232,GA24}.

We have already calculated the unpolarized valence quark distributions of pion using the modified $\chi QM$ \cite{NYVPi}. In this low-energy model applied in the non-perturbative region of QCD, the essential degrees of freedom are constituent quarks, GS bosons and gluon \cite{NYVPi,Chi1,Chi2,Chi3,Chi4,Chi5,Chi6,NYSGPi}. In this work we use the valence quark distributions of pion obtained based on the modified $\chi QM$ to compute the unpolarized valence GPDs of pion. For this purpose, we apply a theoretical model that relates directly the pion's valence GPDs to its standard valence PDFs \cite{PiGPD12}.

The organization of this article is as follows: the relation between valence GPDs and PDFs of pion is described in section \ref{sec-GPDs}. We review the modified $\chi QM$ in subsection \ref{subsec-chiQM}, briefly and obtain the valence quark distributions of pion in this subsection. We present the results of our model for valence GPDs in subsection \ref{subsec-ResVGPDs}. In section \ref{FFs}, we calculate the EMFF and GFF of pion applying the GPDs obtained in section \ref{sec-GPDs} and present the results in this section. We give our conclusions in section \ref{con}.
\section{\label{sec-GPDs}GPDs of Pion}
In this section we calculate the valence generalized parton distributions of pion using a theoretical approach given in Ref. \cite{PiGPD12}. In the first step the GPDs of pion at renormalization scale $\mu$ in terms of the light-front wave functions are written as \cite{PiGPD8,PiGPD12,PiGPD13}:
\begin{eqnarray}
H_q^{\pi}(x,{\xi},t;\mu)=
\int \frac{d^2k_\perp}{16{\pi}^3} {{\psi}_q^{\pi}}^{\ast}(x_-,k^2_{{\perp}-};\mu){\psi}_q^{\pi}(x_+,k^2_{{\perp}+};\mu),\nonumber\\\label{LFWF}\end{eqnarray}
in which ${\psi}$ is the LFWF; $t=-{\Delta}^2=-(p^{\prime}-p)^2$ with $p$, $p^{\prime}$ being the initial and final momenta of pion in the scattering process, respectively. The skewness $\xi=[-n.\Delta]/[2n.P]$; $P=(p+p^{\prime})/2$; $x_{\pm}$ and $k_{{\perp}\pm}$ are defined as \cite{PiGPD8,PiGPD12,PiGPD13}:
\begin{eqnarray}
x_{\pm}&=&\frac{x\pm\xi}{1\pm\xi},\nonumber\\
k_{{\perp}\pm}&=&k_{\perp}\mp\frac{1-x_{\pm}}{1\pm\xi}\frac{\Delta_{\perp}}{2},\label{xpm}\end{eqnarray}
where $\Delta_{\perp}$ denotes the momentum transfer in the hadron's transverse plane. In Eq.(\ref{LFWF}), the GPDs are defined on $\vert x \vert\geq\xi$ that leads to obtaining them within DGLAP domain. In next step by considering factorized (separable) \textit{Ansatz}, the LFWF for the pair of quark-antiquark $(q^\prime\bar{q^{\prime\prime}})$ with helicity $(\lambda^\prime\lambda^{\prime\prime})$ is regarded as \cite{PiGPD12}:
\begin{equation}
{\psi}_{q^\prime\bar{q^{\prime\prime}}/\lambda^\prime\lambda^{\prime\prime}}^{\pi}(x,k_{\perp}^2;\mu)=
f_{q^\prime\bar{q^{\prime\prime}}}^{\pi}(x;\mu)g_{{q^\prime\bar{q^{\prime\prime}}/\lambda^\prime\lambda^{\prime\prime}}}^{\pi}(k_{\perp}^2;\mu).\label{LFWF1}
\end{equation}
The above LFWF should satisfy the following sum rule \cite{PiGPD12}:
\begin{equation}
q^{\pi}(x;\mu)=\sum_{\substack{
  \lambda^\prime,\lambda^{\prime\prime} \\
  q^\prime,\bar{q^{\prime\prime}}}
  }
\delta_{qq^\prime}\int \frac{d^2k_\perp}{16{\pi}^3}{\vert {\psi}_{q^\prime\bar{q^{\prime\prime}}/\lambda^\prime\lambda^{\prime\prime}}^{\pi}(x,k_{\perp}^2;\mu) \vert}^2,\label{qpi}
\end{equation}
in which the leading-twist valence quark distribution is denoted by $q^{\pi}(x;\mu)$.

In further step, by considering a simple form for $f_{q^\prime\bar{q^{\prime\prime}}}^{\pi}$ as $f_{q^\prime\bar{q^{\prime\prime}}}^{\pi}(x;\mu)=\sqrt{q^{\pi}(x;\mu)}$ that fulfill Eq.(\ref{qpi}) and inserting it into Eq.(\ref{LFWF}), the DGLAP-domain valence GPDs of pion are obtained \cite{PiGPD12}:
\begin{equation}
H_q^{\pi}(x,{\xi},t;\mu)=
\sqrt{q^{\pi}(x_+;\mu)q^{\pi}(x_-;\mu)}~{\Phi}_q^{\pi}(x,{\xi},t;\mu),\label{VGPD}
\end{equation}
where
\begin{align}
&{\Phi}_q^{\pi}(x,{\xi},t;\mu)=\nonumber\\&
\sum_{\substack{
  \lambda^\prime,\lambda^{\prime\prime} \\
  q^\prime,\bar{q^{\prime\prime}}}
  }
\delta_{qq^\prime}\int \frac{d^2k_\perp}{16{\pi}^3}{g_{{q^\prime\bar{q^{\prime\prime}}/\lambda^\prime\lambda^{\prime\prime}}}^{\pi{\ast}}}(k^2_{{\perp}+};\mu){g_{{q^\prime\bar{q^{\prime\prime}}/\lambda^\prime\lambda^{\prime\prime}}}^{\pi}}(k^2_{{\perp}-};\mu).\nonumber\\\label{Phi1}
\end{align}
As can be seen above, the pion GPDs are dependent on momentum transfer via ${\Phi}_q^{\pi}$ function that arises from factorized \textit{Ansatz} for LFWFs.

Finally, the $\Phi$ function is obtained by taking the two independent helicity states of quarks into account \cite{PiGPD12,CMMR}:
\begin{eqnarray}
{\Phi}_q^{\pi}(x,{\xi},t;\mu)&=&\frac{1}{4}\frac{1}{1+{\zeta}^2}\nonumber\\
&&\times\left(3+\frac{1-2\zeta}{1+\zeta}\frac{\textrm{arctanh}\left(\sqrt{\frac{\zeta}{1+\zeta}}\right)}{\sqrt{\frac{\zeta}{1+\zeta}}}\right).\label{PhiFunc}
\end{eqnarray}
In Eq.(\ref{PhiFunc}), $\zeta=\frac{-t}{4M^2}\frac{(1-x)^2}{\left(1-\xi^2\right)}$ and $M$ denotes the mass scale. It should be pointed that at $t=0$, ${\Phi}_q^{\pi}(x,{\xi},0;\mu)=1$ and hence $H_q^{\pi}(x,{\xi},0;\mu)=\sqrt{q^{\pi}(x_+;\mu)q^{\pi}(x_-;\mu)}$ that shows these GPDs satisfy the positivity property at $t=0$ and $H_q^{\pi}(x,0,0;\mu)=q^{\pi}(x;\mu)$ as expected.

One important property of above approach is that in which the valence GPDs of pion are directly related to its valence PDFs (see Eq.(\ref{VGPD})). Indeed the 3D picture of pion can be given using the investigation of its one-dimensional structure. So, in order to obtain the valence GPDs of pion, we should have its valence quark distribution functions. We use the results of modified chiral quark model for these valence distributions that have been calculated in our previous work \cite{NYVPi}. We review the modified $\chi QM$ in subsection \ref{subsec-chiQM}. 
\subsection{\label{subsec-chiQM}The Valence PDFs of Pion in the Modified $\chi QM$ } 
In this subsection we describe the method of calculating the valence quark distributions of $\pi^+$ meson based on the modified chiral quark model, briefly. In this model the structure of pion is considered as a bound state of a bare quark ($u_0$) and a bare antiquark ($\bar{d_0}$) dressed by the clouds of GS bosons \cite{Chi5} and gluon \cite{NYVPi}. The quark distribution function of pion is computed by formulating the interactions which can be occurred in the GS boson (gluon)-bare quark vertex. The direct contribution of GS boson (gluon) dressing correction to the quark distribution appears in the formulation as the following convolution integral \cite{NYVPi,Chi1,Chi2,Chi3,Chi4,Chi5,Chi6,NYSGPi}:
\begin{equation}
q_{j}(x)=P_{j{\cal
B}(g)/i} \otimes q_{0}=\int^{1}_{x}\frac{\textmd{d}y}{y}~P_{j{\cal
B}(g)/i}(y)~q_{0,i}(\frac{x}{y}),\label{q}
\end{equation}
where the distribution of the bare quark $i$ is denoted by $q_{0,i}$ and $P_{j{\cal B}(g)/i}$ is the splitting function that gives the probability of finding a constituent quark $j$ with a GS boson $\cal B$ (gluon $g$) coming from the parent quark $i$ \cite{NYVPi,Chi1,Chi2,Chi3,Chi4,Chi5,Chi6,NYSGPi}. Similar interactions can occur in the vertex of the GS boson (gluon)-bare antiquark that contribute to the antiquark distributions of $\pi^+$ via the corresponding relation of Eq.(\ref{q}).
\begin{figure*}[htp]
  \begin{center}
    \begin{tabular}{c}
      {\includegraphics[width=86mm,height=65mm]{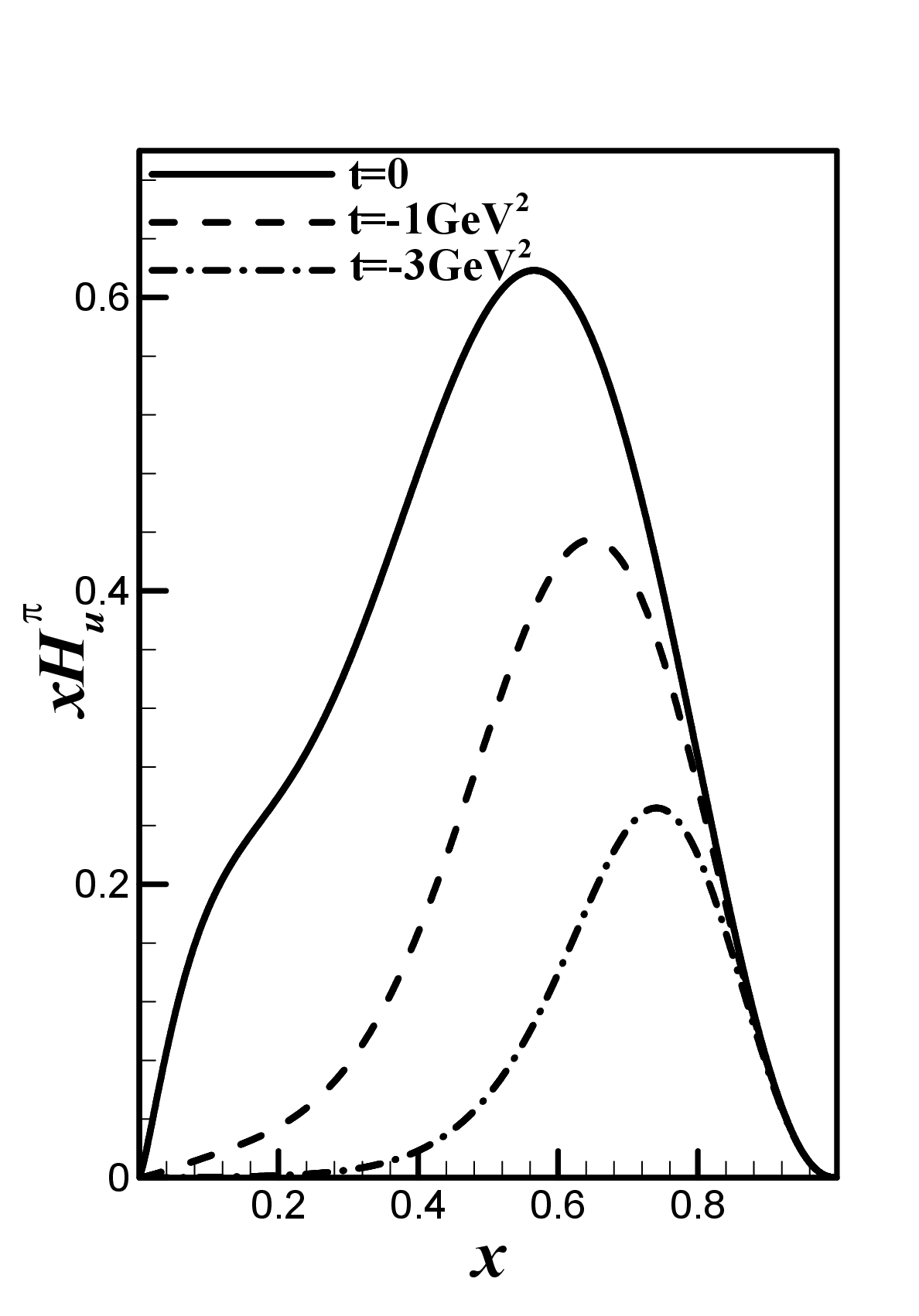}} \\
        \end{tabular}
\caption{ The valence GPDs of pion $xH_u^{\pi}(x,0,t)$ as the functions of $x$ at three fix values of $t$ and the scale $\mu_0^{2}=0.25~\textrm{GeV}^2$. \label{fig:1}}
      \end{center}
\end{figure*}
\begin{figure*}[htp]
  \begin{center}
    \begin{tabular}{c}
      {\includegraphics[width=86mm,height=65mm]{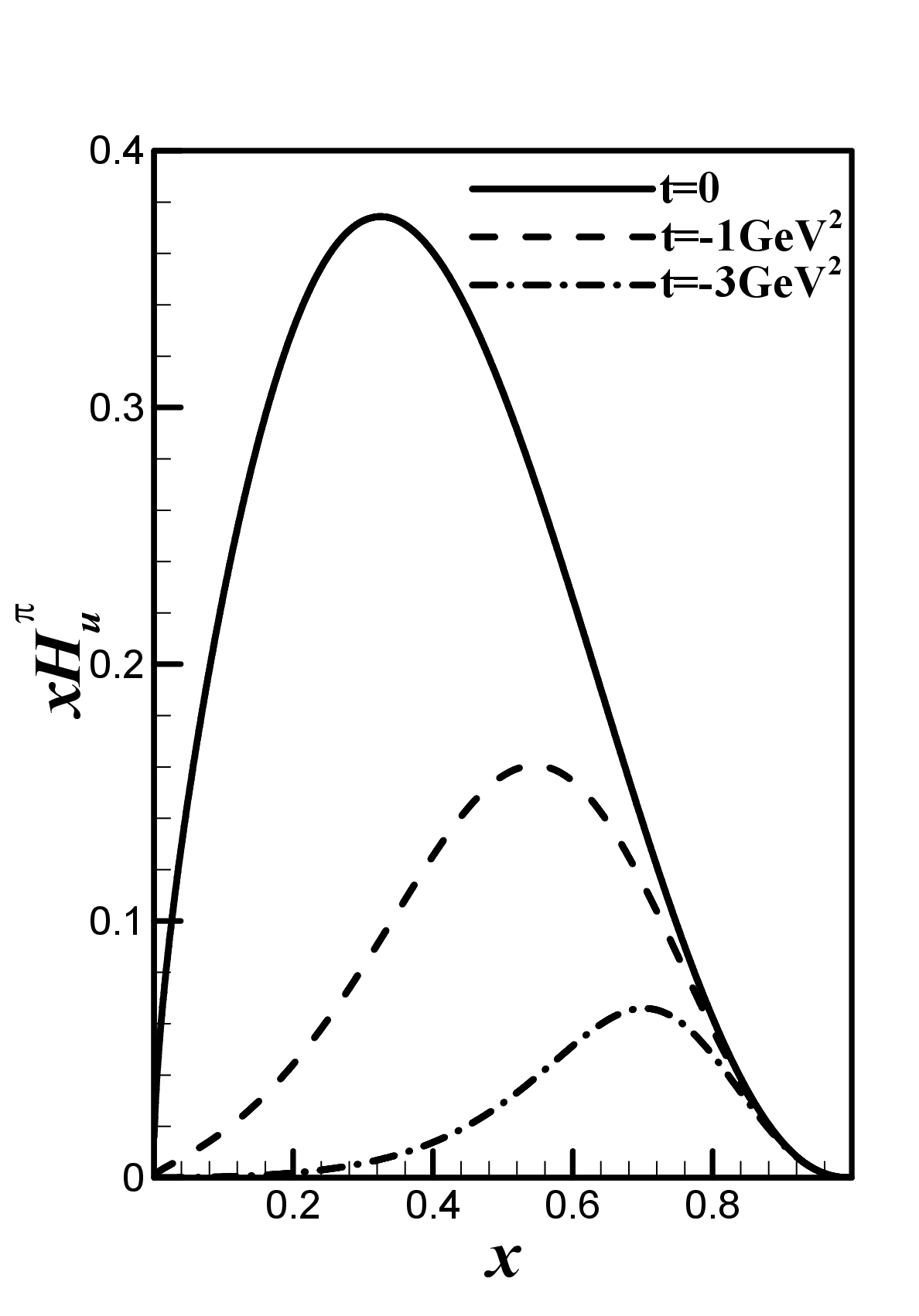}} \\
        \end{tabular}
\caption{ The valence GPDs of pion $xH_u^{\pi}(x,0,t)$ as the functions of $x$ at three fix values of $t$ and the scale $\mu^{2}=4~\textrm{GeV}^2$. \label{fig:2}}
      \end{center}
\end{figure*}

On the other hand, we should regard the contribution of quark-antiquark pair production (emission) by GS boson (gluon) in the quark distribution of pion that is formulated as \cite{NYVPi,Chi1,Chi2,Chi3,Chi4,Chi5,Chi6,NYSGPi}:
\begin{eqnarray}
q_{k}(x)&=&V({\cal P})\otimes P\otimes
q_0\nonumber\\
&&=\int\frac{\textmd{d}y_{1}}{y_{1}}\frac{\textmd{d}
y_{2}}{y_{2}}~V_{k/{\cal B}}({\cal P}_{kg})(\frac{x}{y_{1}})~P_{{\cal B}(g)
j/i}(\frac{y_{1}}{y_{2}})~q_{0,i}(y_{2}).\label{two-split}\nonumber\\
\end{eqnarray}
Here $V_{k/{\cal B}}$ is the probability for finding $q_k$ in the GS boson and ${\cal P}_{kg}$ denotes the standard splitting function of gluon to quarks \cite{NYVPi,Chi4,AP}. The corresponding relation of the above equation for GS boson (gluon)-bare antiquark vertex is written, similarly. 
\begin{figure*}[htp]
  \begin{center}
    \begin{tabular}{c}
      {\includegraphics[width=86mm,height=65mm]{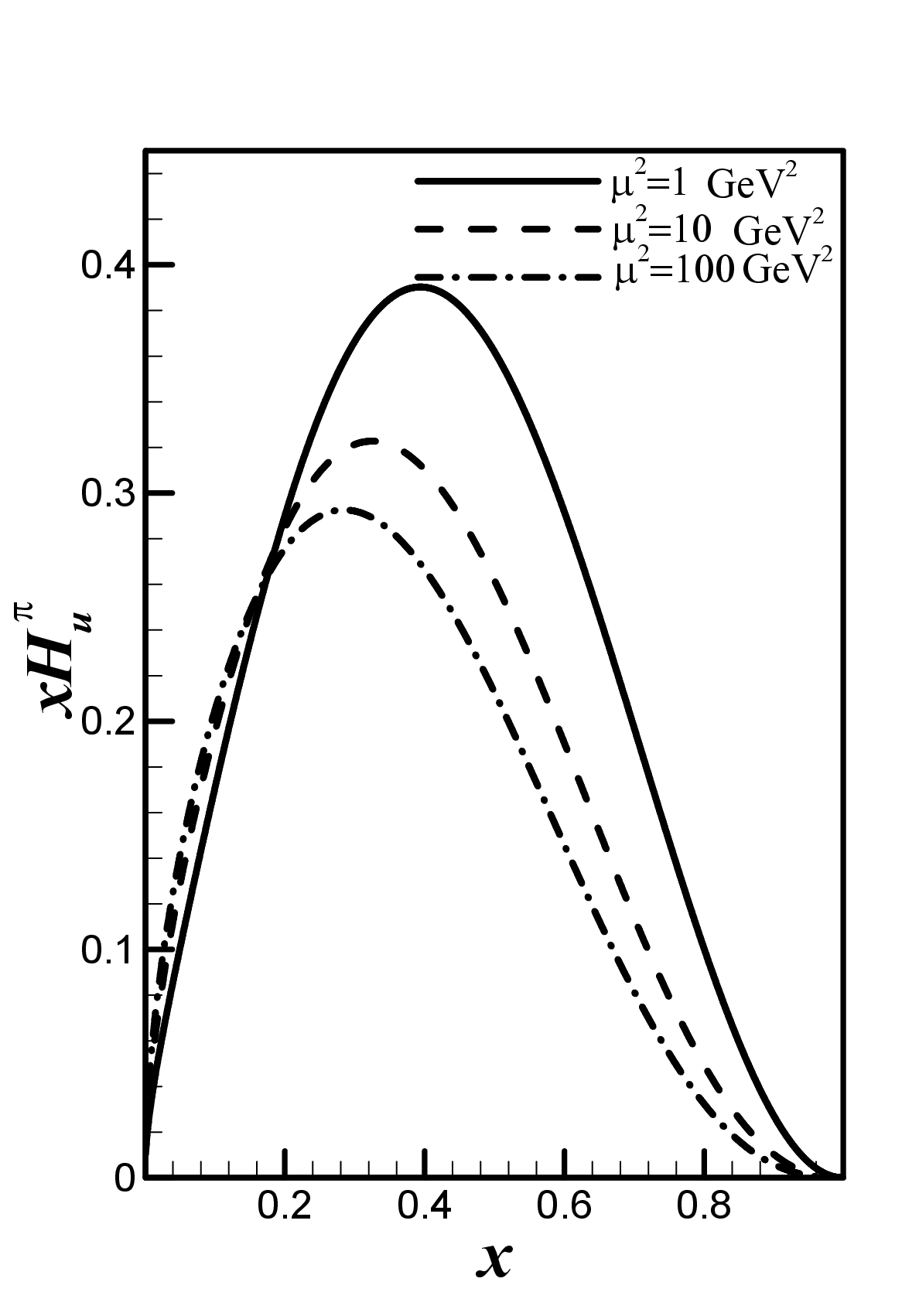}} \\
        \end{tabular}
\caption{ $xH_u^{\pi}(x,0,t)$ with respect to $x$ at three values of $\mu$ and $t=-0.11~\textrm{GeV}^2$. \label{fig:3}}
      \end{center}
\end{figure*}
\begin{figure*}[htp]
  \begin{center}
    \begin{tabular}{ccc}
      {\includegraphics[width=57mm,height=65mm]{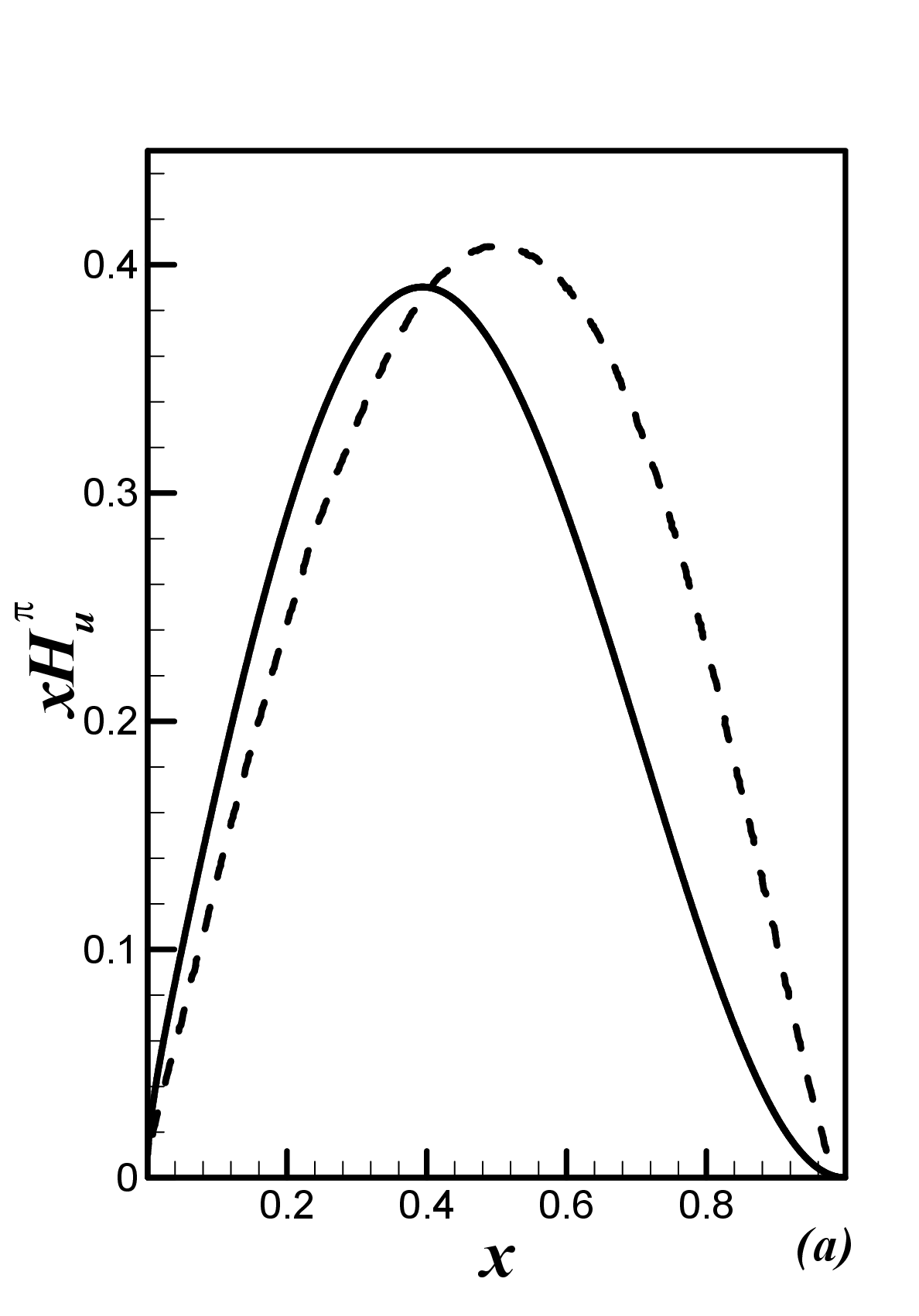}} &
      {\includegraphics[width=57mm,height=65mm]{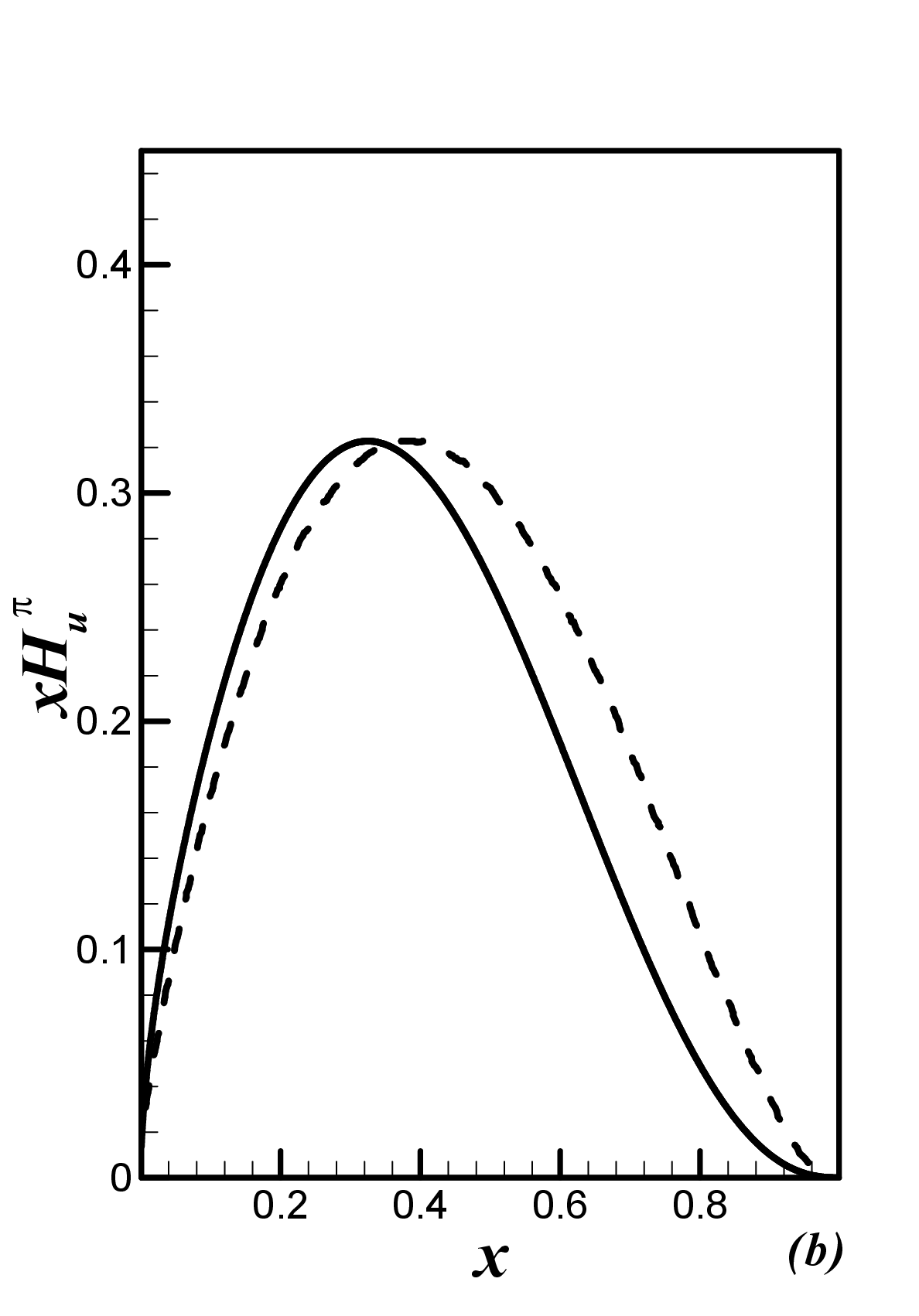}} &
       {\includegraphics[width=57mm,height=65mm]{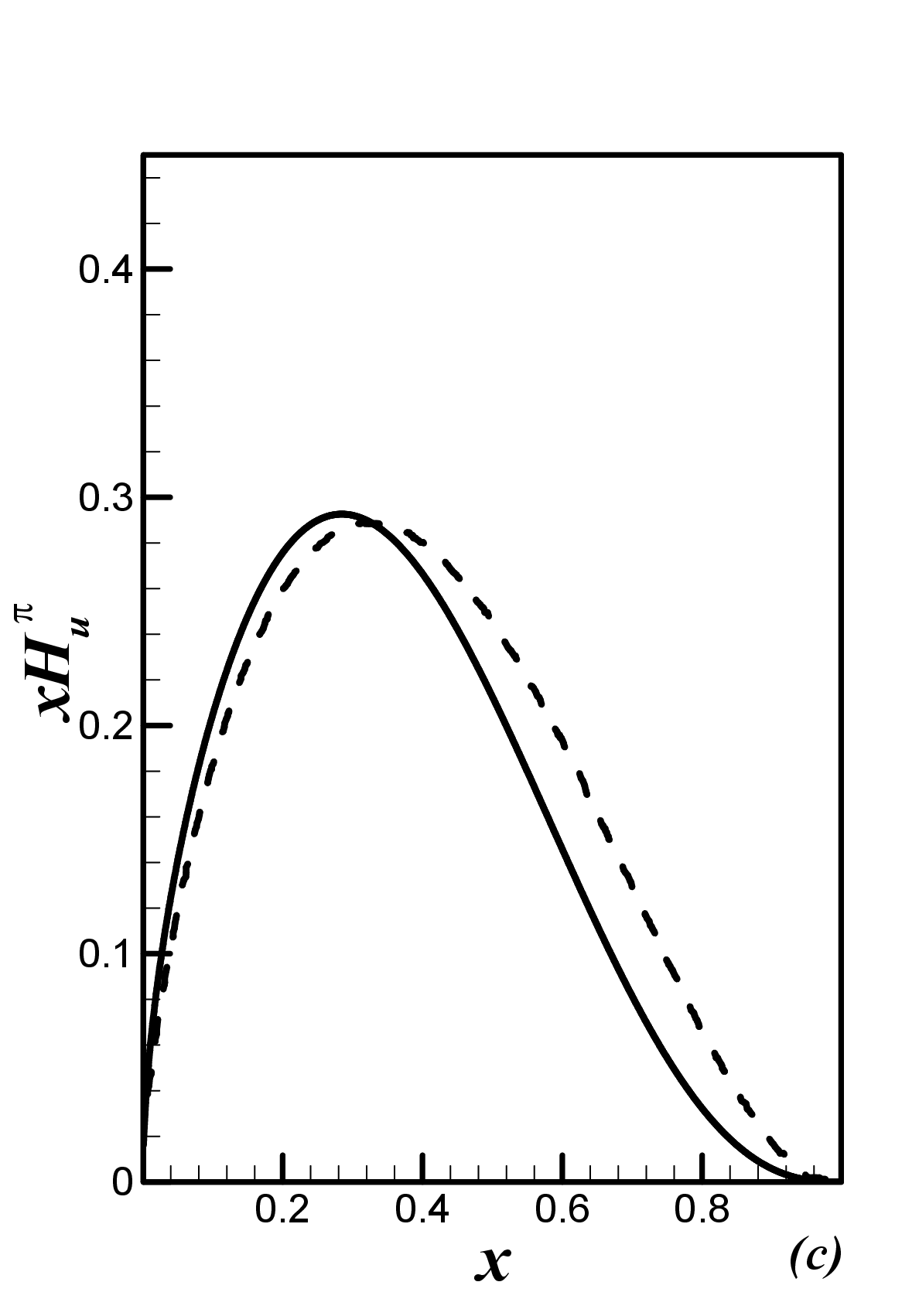}} \\
     \end{tabular}
\caption{ $xH_u^{\pi}(x,0,t)$ with respect to $x$ at $t=-0.11~\textrm{GeV}^2$ and the scale (a) $\mu^{2}=1~\textrm{GeV}^2$, (b) $\mu^{2}=10~\textrm{GeV}^2$ (c) $\mu^{2}=100~\textrm{GeV}^2$. The results of our work (solid lines) are compared with those of Ref. \cite{PiGPD11} (dashed lines).  \label{fig:4}}
      \end{center}
\end{figure*}
So we can achieve $u$ and $\bar{u}$ distributions of $\pi^+$ meson using Eqs.(\ref{q},\ref{two-split})\cite{NYVPi}:
\begin{eqnarray}
u^{\pi}(x)&=&Z^{\pi}_{u}u^{\pi}_{0}(x)+\frac{1}{2}P_{u\pi^{0}/u}\otimes u^{\pi}_{0}+V_{u/\pi^{+}}\otimes P_{\pi^{+}d/u}\otimes u^{\pi}_{0}
\nonumber\\
&&+V_{u/\pi^{+}}\otimes P_{\pi^{+}\bar{u}/\bar{d}}\otimes \bar{d_0}^{\pi}+V_{u/K^{+}}\otimes
P_{K^{+}s/u}\otimes u^{\pi}_{0} \nonumber
\\&&
+\frac{1}{4}V_{u/\pi^{0}}\otimes (P_{\pi^{0}u/u}\otimes
u^{\pi}_{0}+P_{\pi^{0}\bar{d}/\bar{d}}\otimes \bar{d_0}^{\pi}) \nonumber\\&&
+P_{{ug}/{u}}\otimes u^{\pi}_{0}\nonumber\\&&
+\frac{1}{3}{{\cal P}_{ug}}\otimes
(P_{{gu}/{u}}\otimes{u^{\pi}_0}+P_{{g\bar{d}}/{\bar{d}}}\otimes \bar{d_0}^{\pi}),\label{up(x)}
\end{eqnarray}
\begin{eqnarray}
\bar{u}^{\pi}(x)&=& P_{\bar{u}\pi^{+}/\bar{d}}\otimes \bar{d_0}^{\pi}\nonumber\\&&
+\frac{1}{4}V_{\bar{u}/\pi^{0}}\otimes (P_{\pi^{0}u/u}\otimes
u^{\pi}_{0}+P_{\pi^{0}\bar{d}/\bar{d}}\otimes \bar{d_0}^{\pi})\nonumber\\&& +\frac{1}{3}{\cal
P}_{\bar{u}g}\otimes (P_{gu/u}\otimes u^{\pi}_{0}+P_{{g\bar{d}}/{\bar{d}}}\otimes \bar{d_0}^{\pi}).\label{ubarp(x)}
\end{eqnarray}
Finally, the valence quark distribution of pion is obtained as \cite{NYVPi}:
\begin{eqnarray}
u^{\pi}_v(x)&=&u^{\pi}(x)-\bar{u}^{\pi}(x)\nonumber\\&
=&Z^{\pi}_{u}u^{\pi}_{0}(x)+\frac{1}{2}P_{u\pi^{0}/u}\otimes u^{\pi}_{0}+V_{u/\pi^{+}}\otimes P_{\pi^{+}d/u}\otimes u^{\pi}_{0}
\nonumber\\
&&+V_{u/\pi^{+}}\otimes P_{\pi^{+}\bar{u}/\bar{d}}\otimes \bar{d_0}^{\pi}+V_{u/K^{+}}\otimes
P_{K^{+}s/u}\otimes u^{\pi}_{0} \nonumber
\\&&
+P_{{ug}/{u}}\otimes u^{\pi}_{0}-P_{\bar{u}\pi^{+}/\bar{d}}\otimes \bar{d_0}^{\pi}\;,\label{uvp(x)}
\end{eqnarray}
in which $u^{\pi}_{0}$, $\bar{d_0}^{\pi}$ are the bare quark distributions of $\pi^+$ and $Z^{\pi}_{u}$ denotes the renormalization factor of valence quark distribution \cite{NYVPi,Chi5}. We should emphasize that for $\pi^+$ meson $u^{\pi}_v(x)=\bar{d}^{\pi}_v(x)$.

Now by computing the valence quark distributions of pion, we are able to obtain the valence GPDs of this meson applying Eq.(\ref{VGPD}). We should note that from now on we use the notation $H_q^{\pi}(x,\xi,t)\equiv H_q^{\pi}(x,\xi,t;\mu)$ for valence GPDs of pion.
\begin{figure*}[htp]
  \begin{center}
    \begin{tabular}{c}
      {\includegraphics[width=86mm,height=65mm]{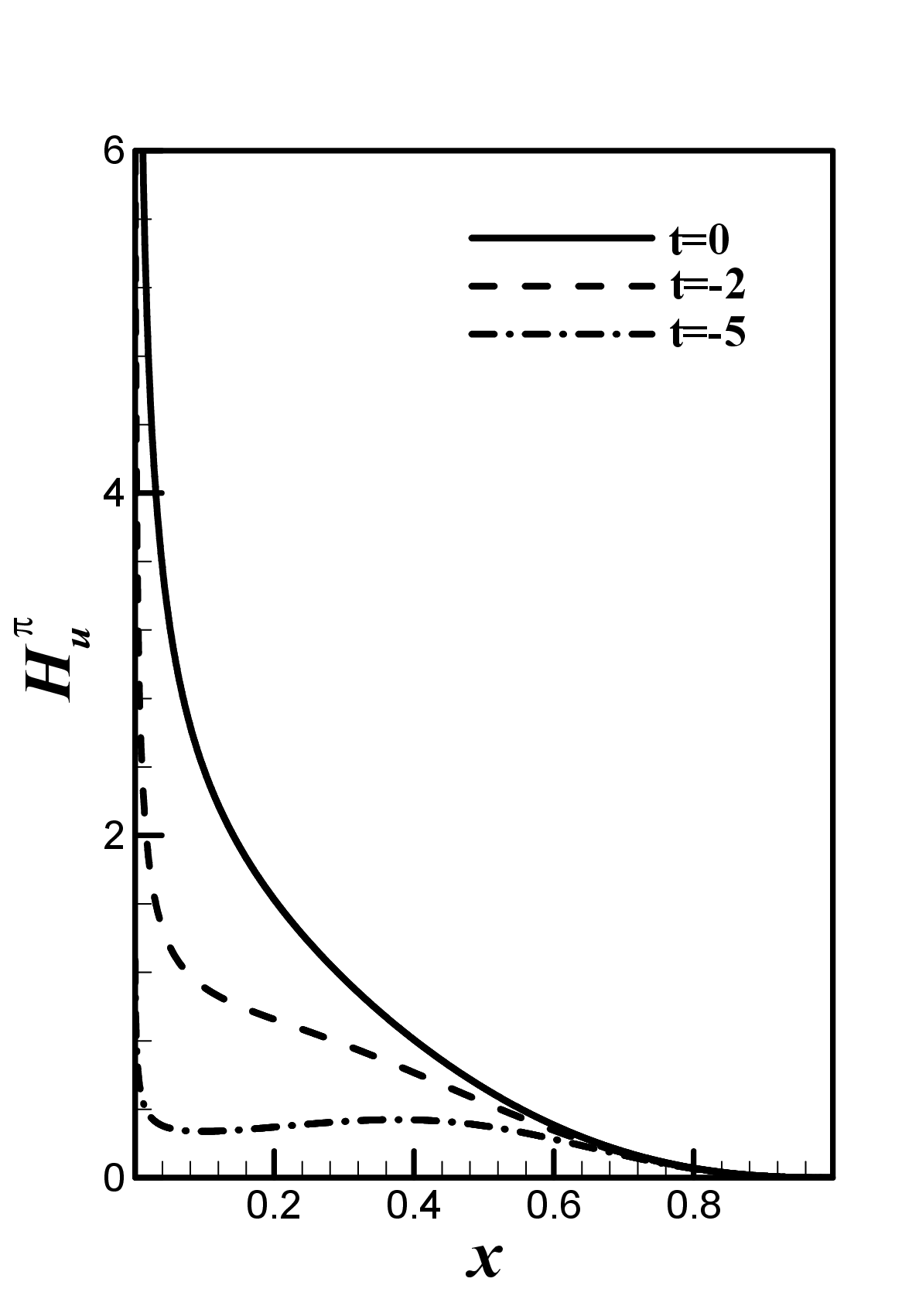}} \\
        \end{tabular}
\caption{ $H_u^{\pi}(x,0,t)$ as a function of $x$ at three fix values of $t\in\lbrace0,-2,-5\rbrace(2\pi/L)^2$ and the scale $\mu^{2}=16~\textrm{GeV}^2$. \label{fig:5}}
      \end{center}
\end{figure*}
\begin{figure*}[htp]
  \begin{center}
    \begin{tabular}{ccc}
      {\includegraphics[width=57mm,height=65mm]{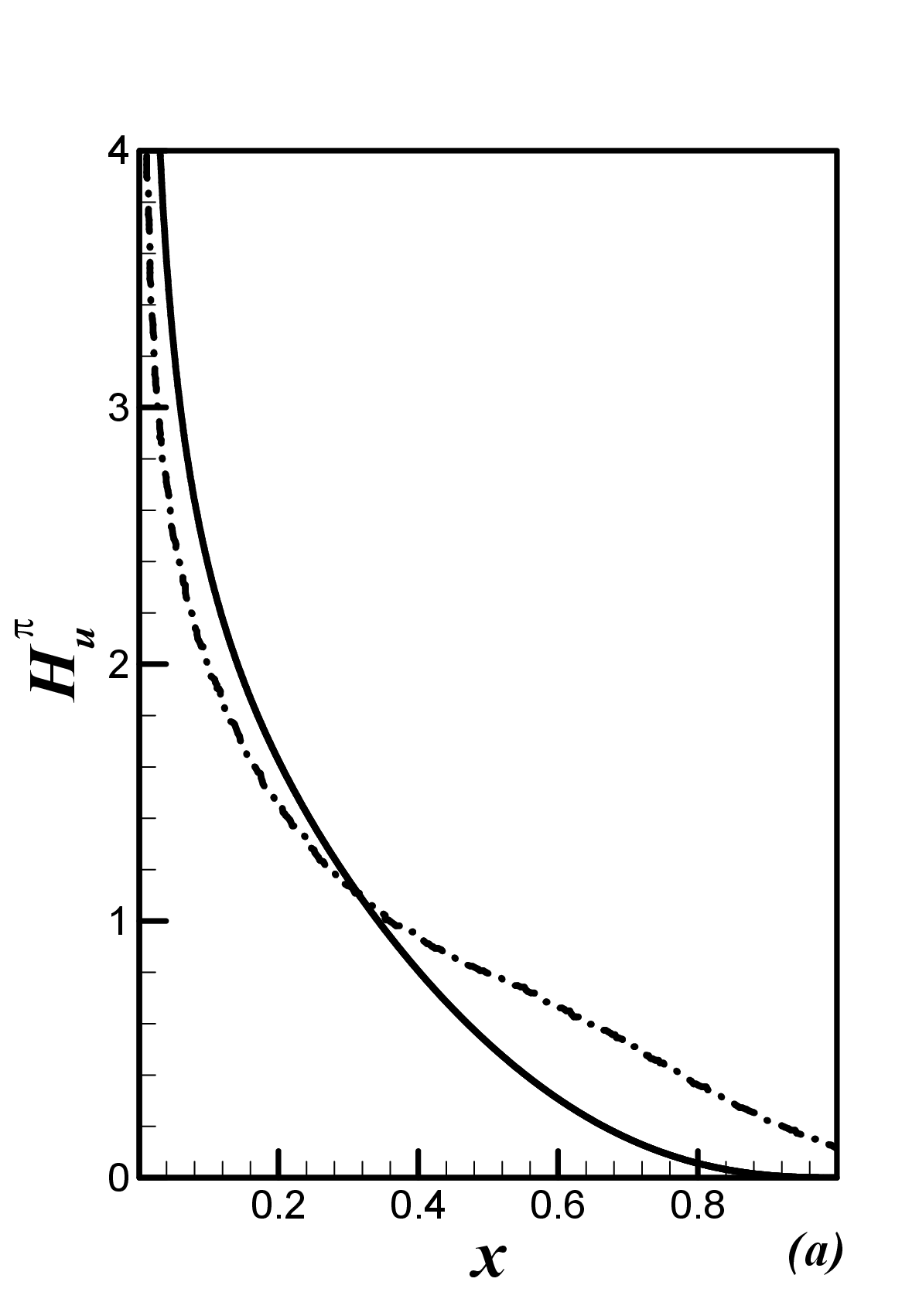}} &
      {\includegraphics[width=57mm,height=65mm]{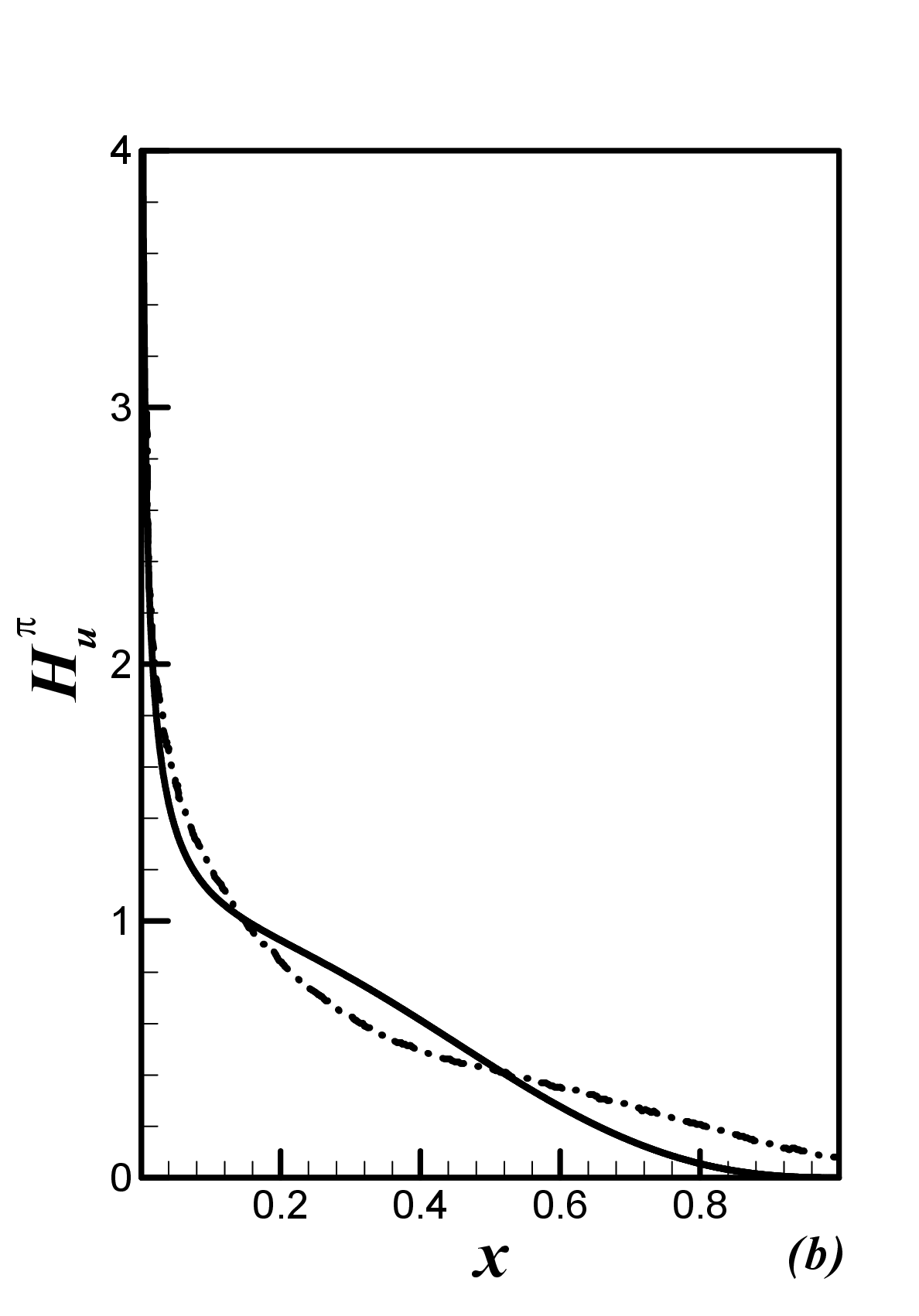}} &
       {\includegraphics[width=57mm,height=65mm]{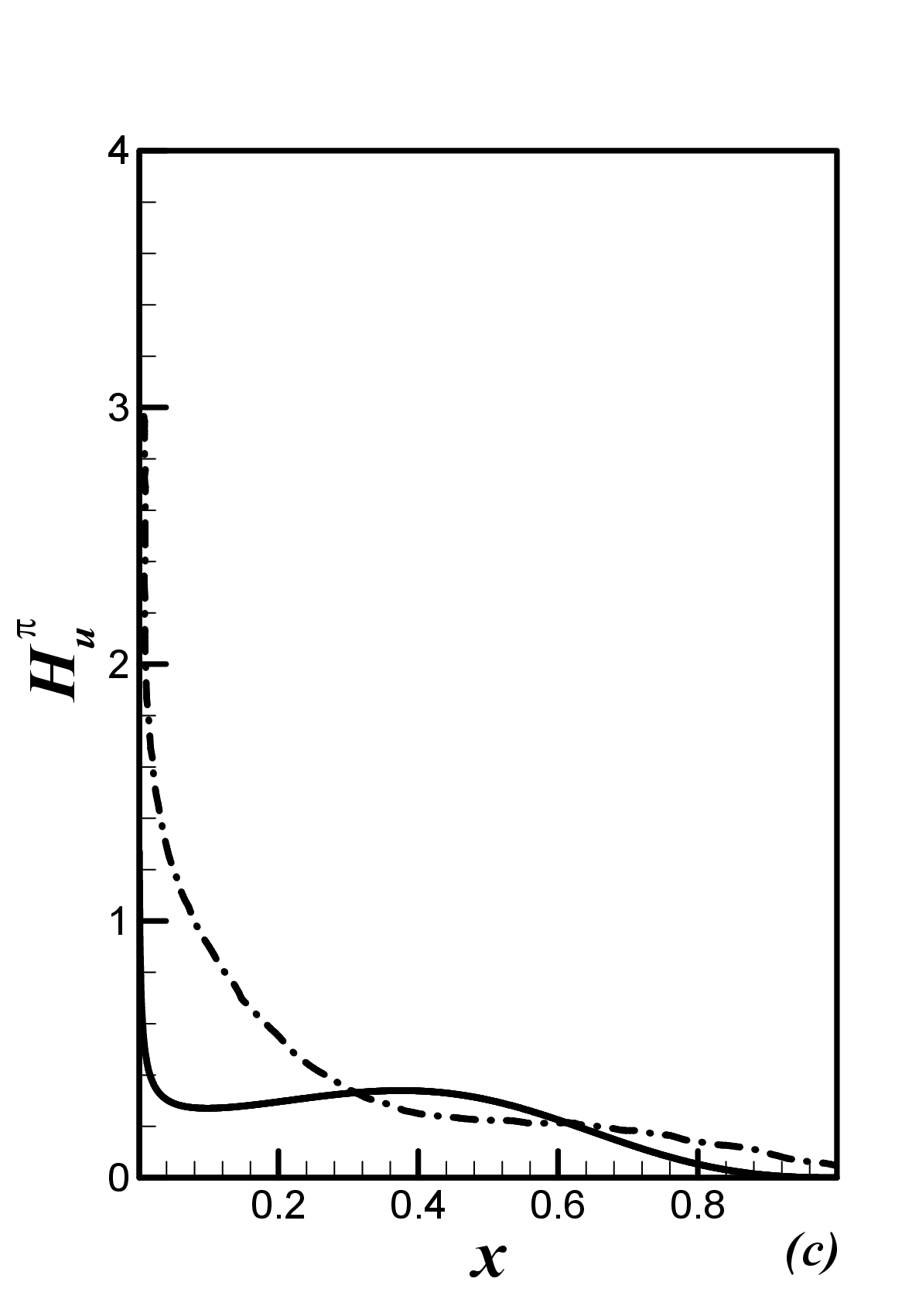}} \\
     \end{tabular}
\caption{ $H_u^{\pi}(x,0,t)$ with respect to $x$ at $\mu^{2}=16~\textrm{GeV}^2$ and the fixed values of $t$: (a) $t=0$, (b) $t=-2(2\pi/L)^2$ (c) $t=-5(2\pi/L)^2$. The results of our work (solid lines) are compared with those of Ref. \cite{PiGPD6} (dashed-dotted lines).  \label{fig:6}}
      \end{center}
\end{figure*}
\subsection{\label{subsec-ResVGPDs}Results of Valence GPDs of Pion}
We compute the DGLAP-domain valence GPDs of $\pi^+$ meson ($H_u^{\pi}$) at zero skewness using Eq.(\ref{VGPD}). To this end, we first obtain the valence quark distribution ($u_v^{\pi}$) applying modified $\chi QM$ at the scale $\mu_0^{2}=0.25~\textrm{GeV}^2$.

We present the results for $xH_u^{\pi}(x,0,t)$ with respect to $x$ at three fix values of $t$; $t=0, t=-1~\textrm{GeV}^2$ and $t=-3~\textrm{GeV}^2$, in Fig.\ref{fig:1}. We should emphasize that $H_u^{\pi}(x,0,0)=u_v^{\pi}(x)$. As can be seen in Fig.\ref{fig:1}, by increasing the $-t$ value, the position of the peak of $xH_u^{\pi}$ shifts toward $x=1$ as expected. In Fig.\ref{fig:2}, we show the same results as Fig.\ref{fig:1} at the scale $\mu^{2}=4~\textrm{GeV}^2$. For this purpose, we first evolve the valence GPDs from $\mu_0^{2}=0.25~\textrm{GeV}^2$ to $\mu^{2}=4~\textrm{GeV}^2$ using DGLAP evolution equations \cite{Chi4,AP}. It is shown that in this higher scale the peak of $xH_u^{\pi}$ at $t=0$ is located at smaller value of $x$ in comparison with the same one in Fig.\ref{fig:1}. In fact by increasing the $\mu$ value, the contribution of sea quarks and gluon in the structure of pion appear more significantly and it affects on the result of valence GPDs.

We also evolve the valence GPDs of pion to the scales $\mu^{2}=1~\textrm{GeV}^2$, $\mu^{2}=10~\textrm{GeV}^2$ and $\mu^{2}=100~\textrm{GeV}^2$ and depict the results for $xH_u^{\pi}(x,0,t)$ as the functions of $x$ at $t=-0.11~\textrm{GeV}^2$ in Fig.\ref{fig:3}. These results at each value of $\mu$ are compared with the results of Ref. \cite{PiGPD11} in figures \ref{fig:4}a, \ref{fig:4}b and \ref{fig:4}c. From Fig.\ref{fig:3}, we observe that by evolving the valence GPDs to the higher scales, their peaks move to lower $x$ values. As can be seen in Fig.\ref{fig:4}, the results of our model show good agreement with those of Ref. \cite{PiGPD11}.

Finally, the scale evolution of the valence quark GPDs $H_u^{\pi}(x,0,t)$ at $\mu^{2}=16~\textrm{GeV}^2$ and the fixed values of $t$, $t\in\lbrace0,-2,-5\rbrace(2\pi/L)^2$, with $L=3~\textrm{fm}$, with respect to $x$, are depicted in Fig.\ref{fig:5}. We compare our results with those of Ref.\cite{PiGPD6} in Fig.\ref{fig:6}. The results of our work show appropriate general properties in comparison with the results of Ref.\cite{PiGPD6} especially at $t=0$ and $t=-2$.
\begin{figure*}[htp]
  \begin{center}
    \begin{tabular}{cc}
      {\includegraphics[width=70mm,height=65mm]{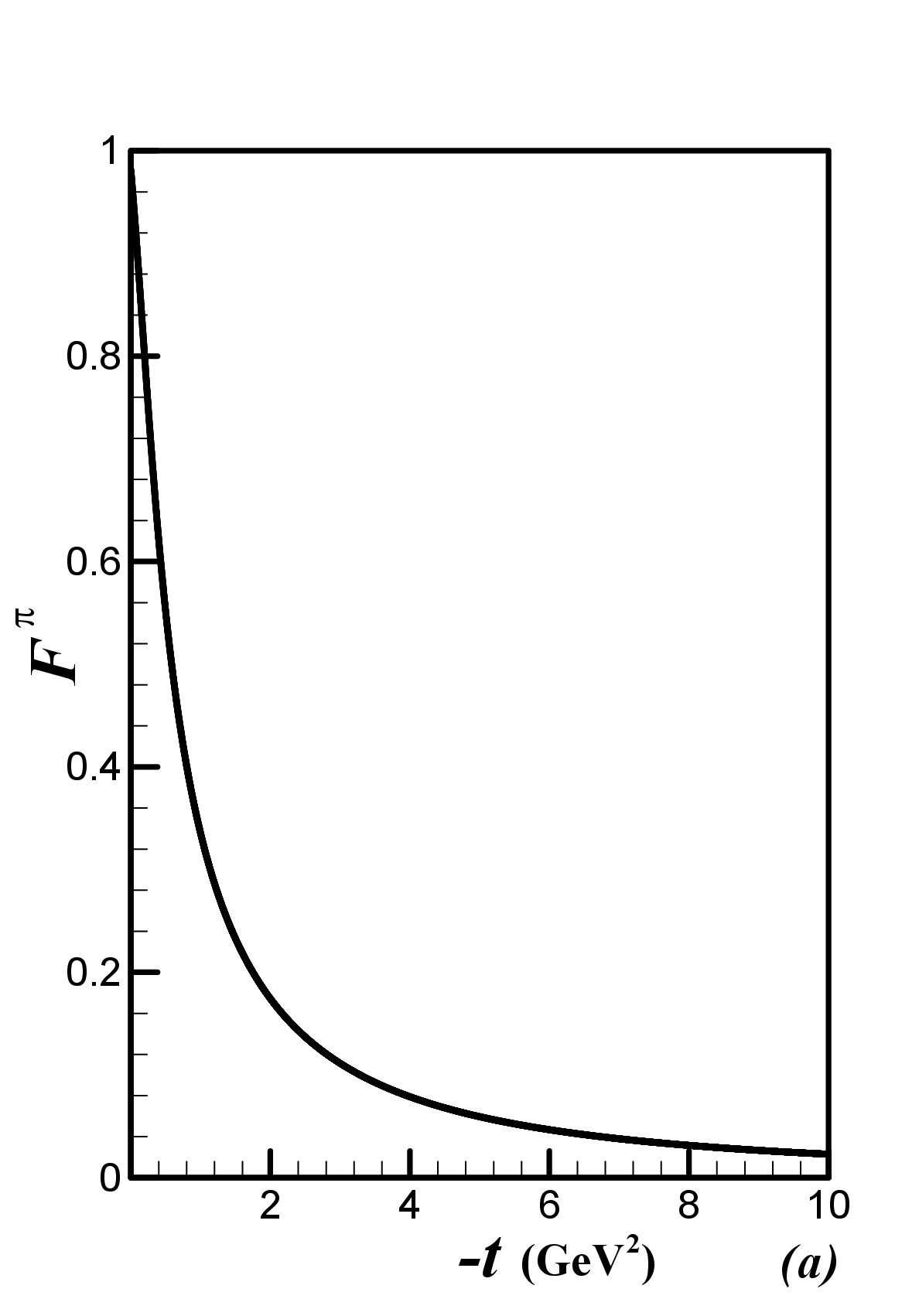}} &
       {\includegraphics[width=70mm,height=65mm]{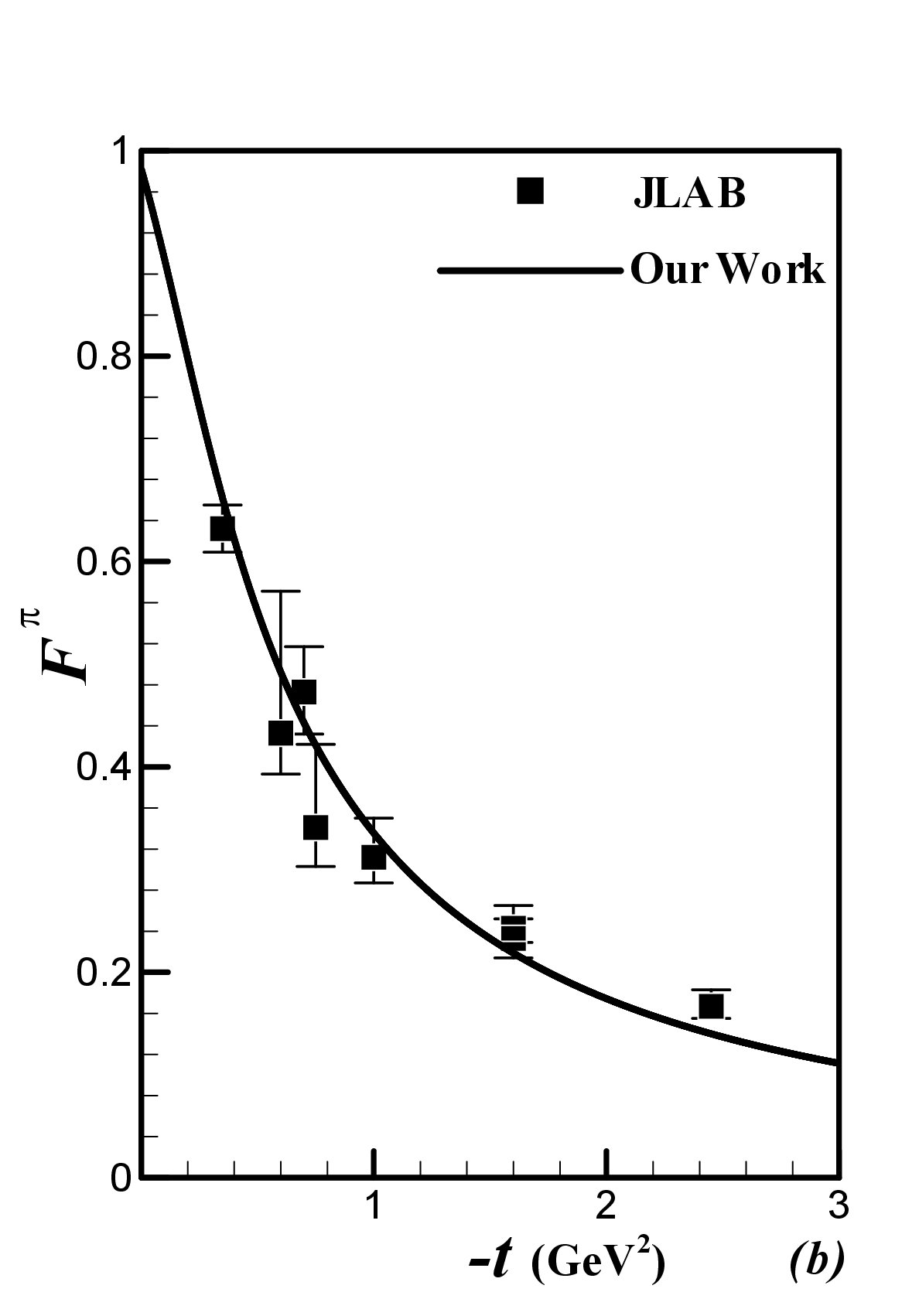}} \\
     \end{tabular}
\caption{ EMFF of pion ($F^{\pi}$) as a function of $-t$. (a) The result of our work for $F^{\pi}$. (b) The result of our work for $F^{\pi}$ in comparison with JLAB experimental data \cite{JLAB}.  \label{fig:7}}
      \end{center}
\end{figure*}
\begin{figure*}[htp]
  \begin{center}
    \begin{tabular}{c}
      {\includegraphics[width=86mm,height=65mm]{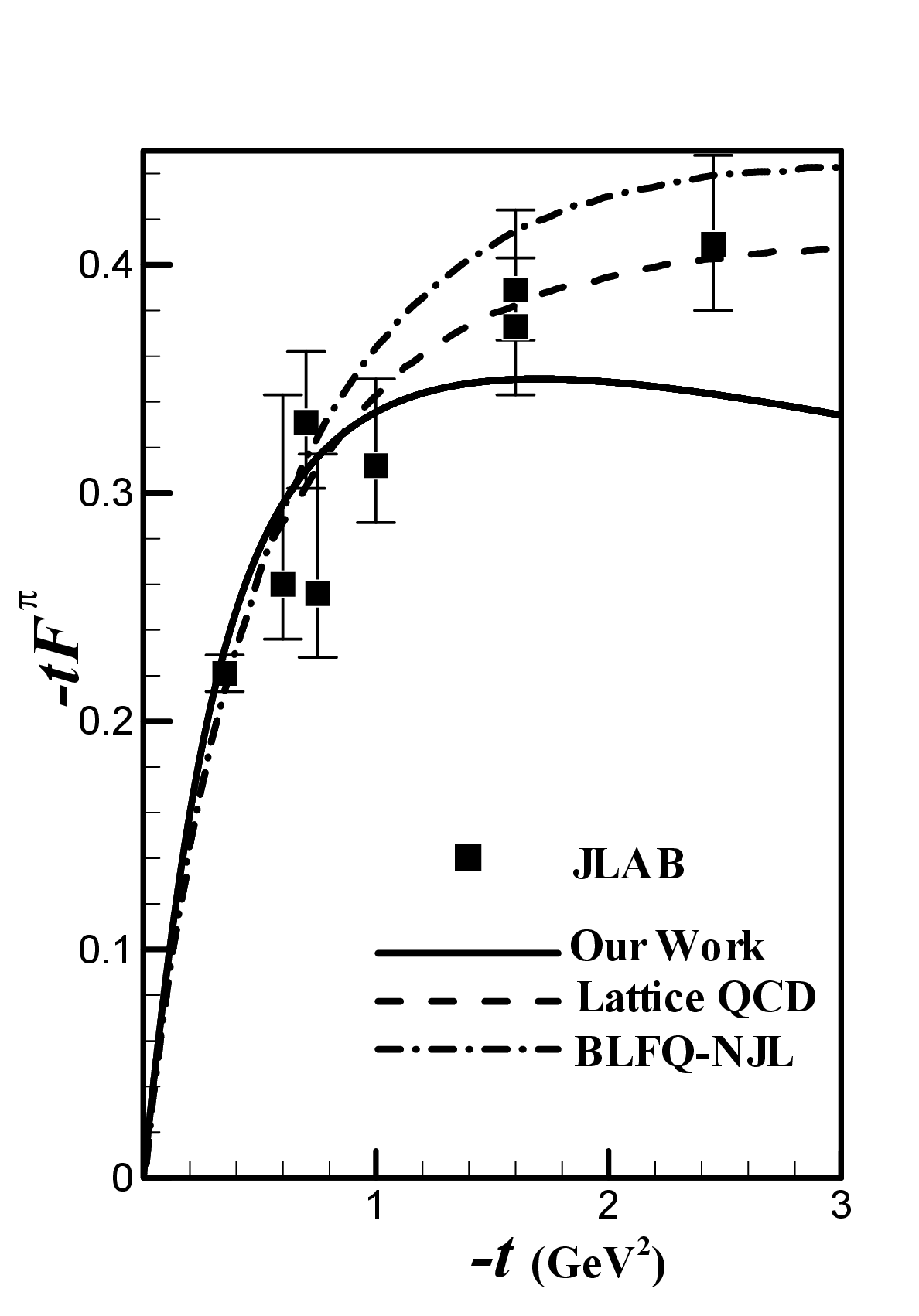}} \\
        \end{tabular}
\caption{ $-tF^{\pi}$ as a function of $-t$. The result of our work is compared with JLAB data \cite{JLAB} and the results of BLFQ-NJL model \cite{PiGPD11} and lattice QCD \cite{LQCD2,PiGPD11}. \label{fig:8}}
      \end{center}
\end{figure*}    
\section{\label{FFs}Electromagnetic and Gravitational Form Factors of Pion}
\begin{figure*}[htp]
  \begin{center}
    \begin{tabular}{cc}
      {\includegraphics[width=70mm,height=65mm]{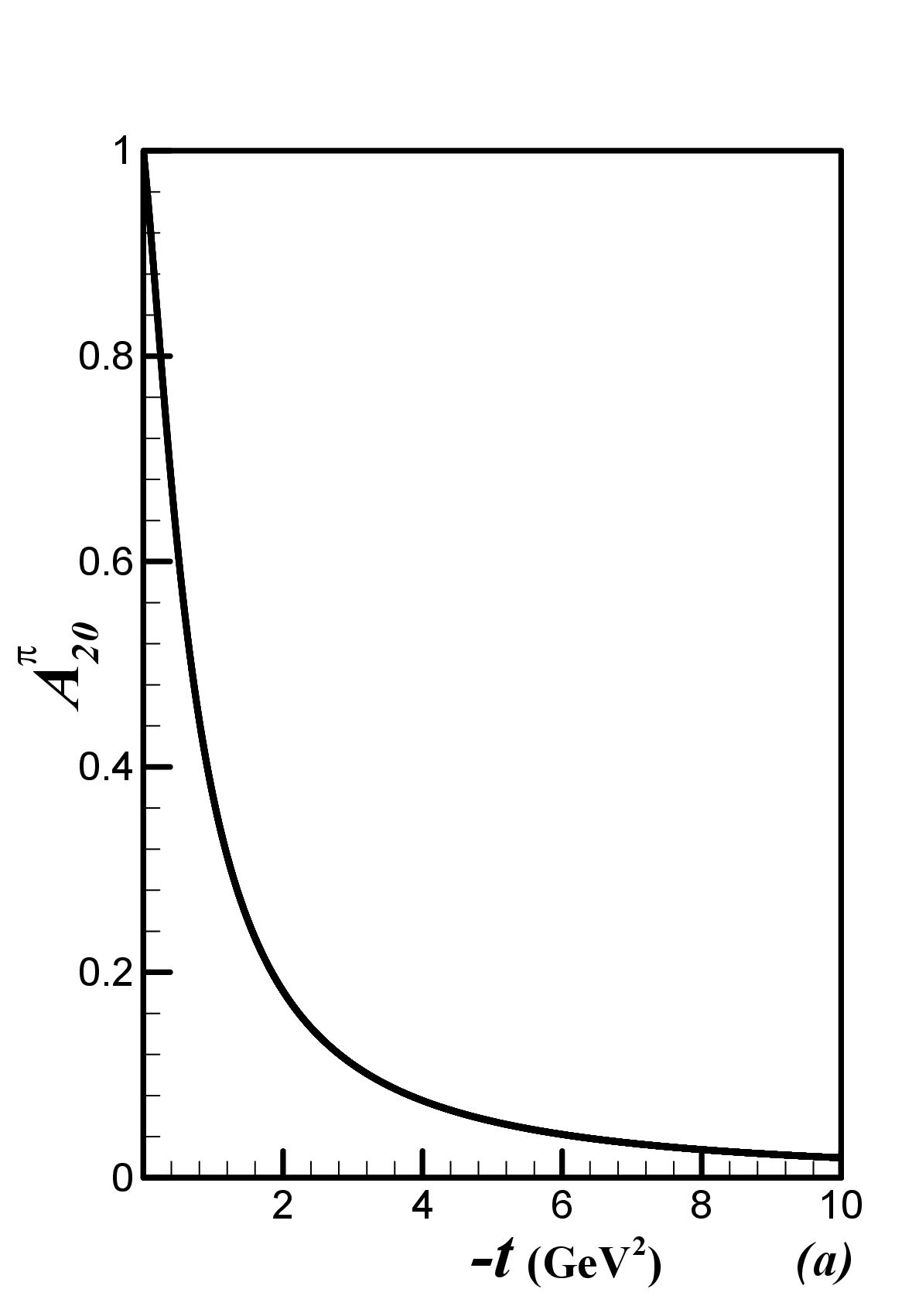}} &
       {\includegraphics[width=70mm,height=65mm]{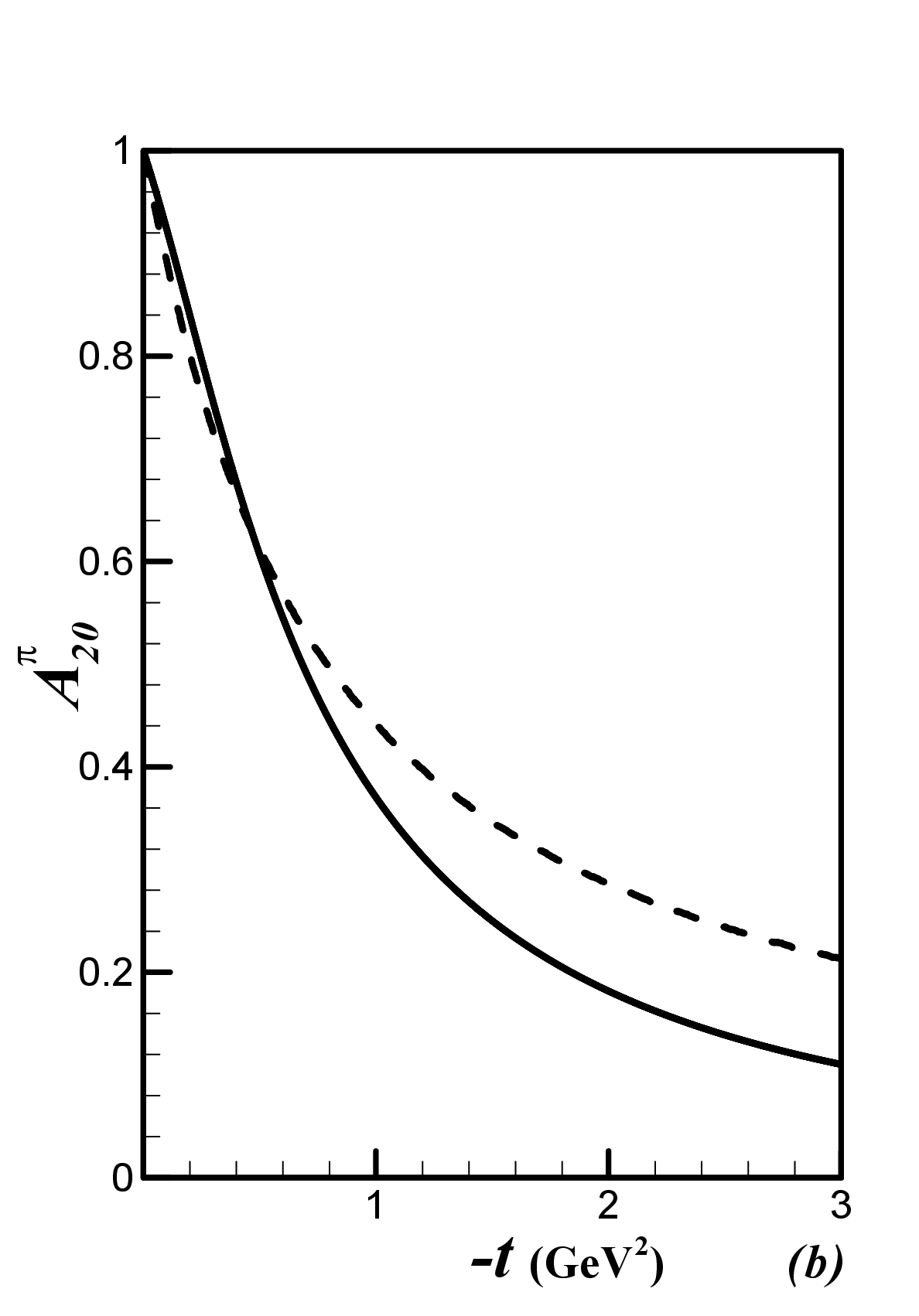}} \\
     \end{tabular}
\caption{ GFF of pion ($A_{20}^{\pi}$) as a function of $-t$ at $\mu^{2}=4~\textrm{GeV}^2$. (a) The result of our work for $A_{20}^{\pi}$. (b) The result of our work for $A_{20}^{\pi}$ (solid line) in comparison with the result of Ref.\cite{PiGPD13} (dashed line).  \label{fig:9}}
      \end{center}
\end{figure*}
\begin{figure*}[htp]
  \begin{center}
    \begin{tabular}{c}
      {\includegraphics[width=86mm,height=65mm]{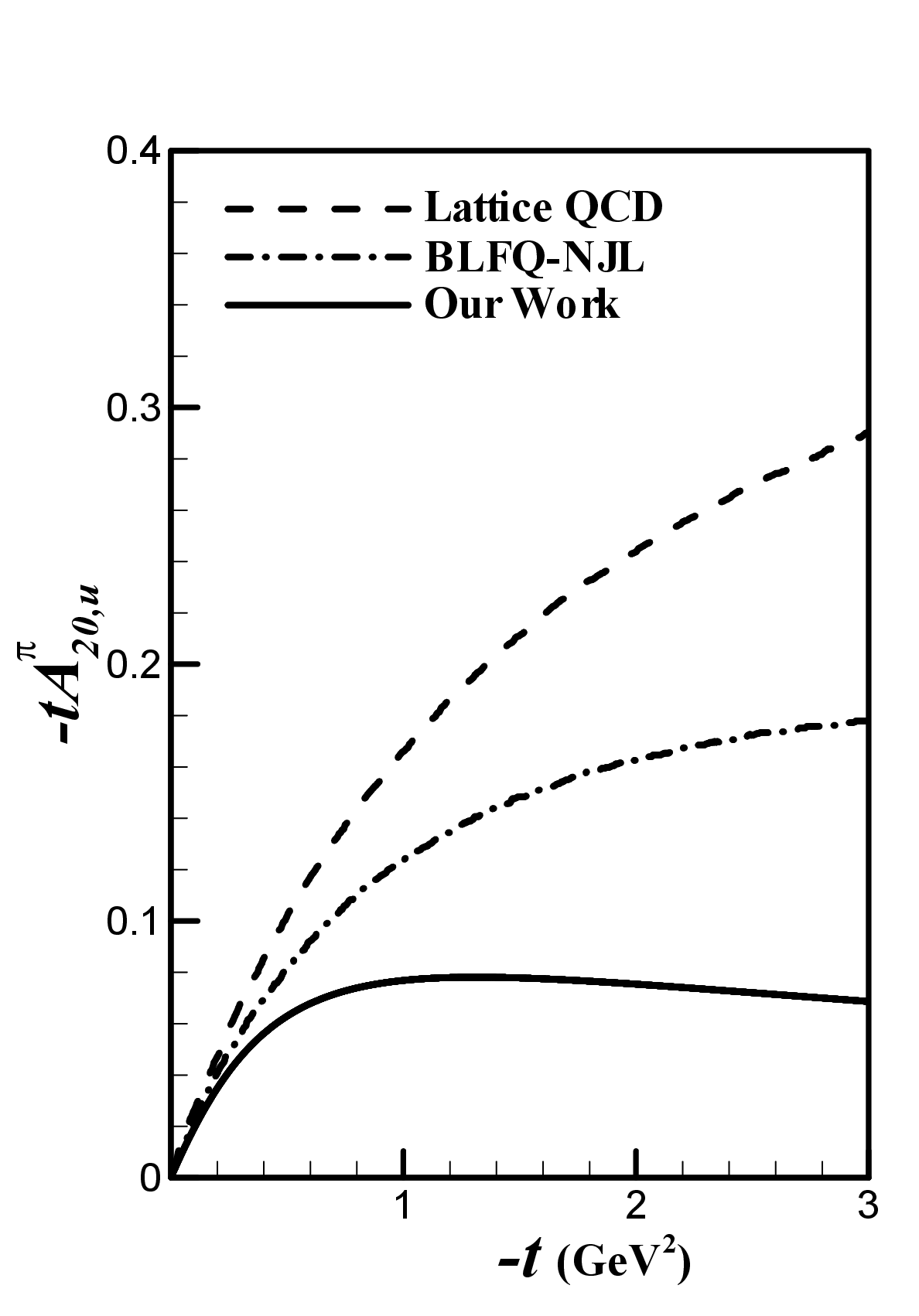}} \\
        \end{tabular}
\caption{ $-tA_{20,u}^{\pi}$ as a function of $-t$ at $\mu^{2}=4~\textrm{GeV}^2$. The result of our work is compared with those of BLFQ-NJL model \cite{PiGPD11} and lattice QCD \cite{LQCD2,PiGPD11}. \label{fig:10}}
      \end{center}
\end{figure*}
In this section we probe the electromagnetic and gravitational FFs of pion using the valence GPDs obtained in section \ref{sec-GPDs}. The contribution of a valence quark to the pion electromagnetic form factor $\left(F_q^{\pi}(t)\right)$ is given through the zeroth moment of the valence quark's GPDs at zero skewness ($\xi=0$) \cite{PiGPD7,PiGPD11,PiGPD12,PiGPD13}:
\begin{equation}
F_q^{\pi}(t)=\int_{-1}^{1} dx~H_q^{\pi}(x,0,t).\label{EFF}
\end{equation}
The complete pion EMFF can be written as \cite{PiGPD7,PiGPD11,PiGPD12,PiGPD13}:
\begin{equation}
F^{\pi}(t)=e_u~F_u^{\pi}(t)+e_{\bar{d}}~F_{\bar{d}}^{\pi}(t),\label{EFFPi}
\end{equation}
in which $e_q$ is the charge of valence quark. It should be noted that the quark EMFF in forward limit, $F_q(0)$ gives the number of valence quarks \cite{PiGPD11}.

We calculate the EMFF of pion using Eq.(\ref{EFFPi}) and present the result in Fig.\ref{fig:7}a. The result of our work for this form factor is depicted in Fig.\ref{fig:7}b in comparison with the JLAB experimental data \cite{JLAB}. It is observed that our result is in good agreement with available experimental data. We also present the result of $-tF^{\pi}(t)$ in Fig.\ref{fig:8} and compare it with the JLAB data \cite{JLAB} and the results of BLFQ-NJL model \cite{PiGPD11} and lattice QCD simulations \cite{LQCD2,PiGPD11}. From Fig.\ref{fig:8} it is found that our result is very close with those of BLFQ-NJL model and lattice QCD within $-t\leqslant1~\textrm{GeV}^2$. As $-t$ increases, the result of our work deviates from the results of above models.

In the further step of our calculations, we compute the charge radius of pion. This radius is defined by \cite{PiGPD7}:
\begin{equation}
r_{\pi}^2=-6\frac{dF^{\pi}(t)}{dt}\mid_{t=0}.\label{PiRad}
\end{equation} 
We obtain $r_{\pi}\simeq0.613~\textrm{fm}$ which is in good agreement with experimental value $r_{\pi}\simeq0.659~\textrm{fm}$ and $r_{\pi}\simeq0.640~\textrm{fm}$ \cite{PiGPD13,PiRad1,PiRad2}.

In the last step of our calculations, we investigate the gravitational form factor of pion. The first-order Mellin moment of valence GPDs gives this GFF as:
\begin{equation}
A_{20,q}^{\pi}(t)=\int_{-1}^{1} dx~xH_q^{\pi}(x,0,t).\label{PiGFF}
\end{equation} 
We calculate the complete GFF of pion $\left(A_{20}^{\pi}(t)\right)$ at the scale $\mu^{2}=4~\textrm{GeV}^2$ applying its evolved valence GPDs and depicte the result in Fig.\ref{fig:9}a. We compare our result for $A_{20}^{\pi}$ with that of Ref.\cite{PiGPD13} in Fig.\ref{fig:9}b. It is observed that the result of our work is close to the result of Ref.\cite{PiGPD13}. The result of our work for $-tA_{20,u}^{\pi}$ as a function of $-t$ is presented in Fig.\ref{fig:10} in comparison with the results of BLFQ-NJL model \cite{PiGPD11} and lattice QCD \cite{LQCD2,PiGPD11} at $\mu^{2}=4~\textrm{GeV}^2$. In this figure essential difference is observed between the results of two above models, while their EMFFs were close to each other (please see Fig.\ref{fig:8}), and also between our result and those of these models especially as $-t$ increases. We should emphasize that, by considering this fact that our results are arisen from a completely theoretical approach, they show appropriate general properties in comparison with the results of other models.    
\section{\label{con}Conclusion}
In conclusion, we have studied the structure of pion based on the modified chiral quark model applied in the non-perturbative region of QCD. We already obtained the valence quark distributions of pion using this model at the scale $\mu_{0}^{2}=0.25~\textrm{GeV}^2$ in our previous work \cite{NYVPi}. Considering these valence distributions, we have calculated the DGLAP-domain valence GPDs of pion at $\mu_{0}^{2}=0.25~\textrm{GeV}^2$ applying a theoretical approach which relates the valence PDFs to valence GPDs \cite{PiGPD12}. We have also evolved the valence GPDs to higher scales using DGLAP evolution equations and compared the obtained results with the results of other works. This is the first time that the valence GPDs of pion are calculated directly from its valence PDFs in the chiral quark model framework.     

In second step of our calculations, we have obtained the electromagnetic and gravitational FFs of pion and presented the results in comparison with available experimental data and the results of other models. It should be noted that we have used a completely theoretical framework in this work. So we can say that the results of our work show acceptable properties comparing the results of other models.

We hope to investigate the GPDs and FFs of kaon and report the results in future.
%
%
\section*{Acknowledgments}
H. Nematollahi would like to thank Physics Department of the University of Tehran for their support and also M. M. Yazdanpanah for interesting discussion and useful comments.

\appendix


\begin{thebibliography}{92}
\bibitem{GPD1} D. M{\"u}ller et al.,
\href{https://doi.org/10.1002/prop.2190420202} {Fortsch. Phys. {\bf 42}, 101 (1994)}.
\bibitem{GPD2} A. V. Radyushkin, 
\href{https://doi.org/10.1016/0370-2693(96)00844-1} {Phys. Lett. B {\bf 385}, 333 (1996)}.
\bibitem{GPD3} A. V. Radyushkin, 
\href{https://doi.org/10.1103/PhysRevD.56.5524} {Phys. Rev. D {\bf 56}, 5524 (1997)}.
\bibitem{GPD4} X. D. Ji, 
\href{https://doi.org/10.1103/PhysRevLett.78.610} {Phys. Rev. Lett. {\bf 78}, 610 (1997)}.
\bibitem{GPD5} X. D. Ji, 
\href{https://doi.org/10.1088/0954-3899/24/7/002} {J. Phys. G {\bf 24}, 1181 (1998)}.
\bibitem{NJL1} W. Broniowski, A. E. Dorokhov and E. Ruiz Arriola, 
\href{https://doi.org/10.1103/PhysRevD.82.094001} {Phys. Rev. D {\bf 82}, 094001 (2010)}.
\bibitem{NJL2} A. E. Dorokhov, W. Broniowski  and E. Ruiz Arriola, 
\href{https://doi.org/10.1103/PhysRevD.84.074015} {Phys. Rev. D {\bf 84}, 074015 (2011)}.
\bibitem{NJL3} A. Freese and I. C. Clo{\"e}t, 
\href{https://doi.org/10.1103/PhysRevC.100.015201} {Phys. Rev. C {\bf 100}, 015201 (2019)}.[Erratum: \href{https://doi.org/10.1103/PhysRevC.105.059901} {Phys. Rev. C {\bf 105}, 059901 (2022)}]
\bibitem{LQCD1} D. Br{\"o}mmel et al. (QCDSF/UKQCD Collaborations), 
\href{https://doi.org/10.1103/PhysRevLett.101.122001} {Phys. Rev. Lett. {\bf 101}, 122001 (2008)}.
\bibitem{LQCD2} D. Br{\"o}mmel et al. (QCDSF/UKQCD Collaborations), 
\href{https://doi.org/10.1140/epjc/s10052-007-0295-6} {Eur. Phys. J. C {\bf 51}, 335 (2007)}.
\bibitem{PiGPD1} M. V. Polyakov and C. Weiss, 
\href{https://doi.org/10.1103/PhysRevD.60.114017} {Phys. Rev. D {\bf 60}, 114017 (1999)}.
\bibitem{PiGPD2} W. Broniowski, E. Ruiz Arriola and K. Golec-Biernat, 
\href{https://doi.org/10.1103/PhysRevD.77.034023} {Phys. Rev. D {\bf 77}, 034023 (2008)}.
\bibitem{PiGPD3} C. Mezrag et al., 
\href{https://doi.org/10.1016/j.physletb.2014.12.027} {Phys. Lett. B {\bf 741}, 190 (2015)}.
\bibitem{PiGPD4} C. Fanelli et al., 
\href{https://doi.org/10.1140/epjc/s10052-016-4101-1} {Eur. Phys. J. C {\bf 76}, 253 (2016)}.
\bibitem{PiGPD5} G. F. de Teramond et al. (HLFHS Collaboration), 
\href{https://doi.org/10.1103/PhysRevLett.120.182001} {Phys. Rev. Lett {\bf 120}, 182001 (2018)}.
\bibitem{PiGPD6} J. W. Chen, H. W. Lin and J. H. Zhang, 
\href{https://doi.org/10.1016/j.nuclphysb.2020.114940} {Nucl. Phys. B. {\bf 952}, 114940 (2020)}.
\bibitem{PiGPD7} C. Shi, K. Bednar, I. C. Clo{\"e}t and A. Freese, 
\href{https://doi.org/10.1103/PhysRevD.101.074014} {Phys. Rev. D {\bf 101}, 074014 (2020)}. 
\bibitem{PiGPD8} J. L. Zhang et al., 
\href{https://doi.org/10.1016/j.physletb.2021.136158} {Phys. Lett. B {\bf 815}, 136158 (2021)}.
\bibitem{PiGPD9} J. L. Zhang, Z. F. Cui, J, Ping and C. D. Roberts,
\href{https://doi.org/10.1140/epjc/s10052-020-08791-1} {Eur. Phys. J. C {\bf 81}, 6 (2021)}.
\bibitem{PiGPD10} C. D. Roberts, D. G. Richards, T. Horn and L. Chang, 
\href{https://doi.org/10.1016/j.ppnp.2021.103883} {Prog. Part. Nucl. Phys. {\bf 120}, 103883 (2021)}.
\bibitem{PiGPD11} L. Adhikari et al. (BLFQ Collaboration),
\href{https://doi.org/10.1103/PhysRevD.104.114019} {Phys. Rev. D {\bf 104}, 114019 (2021)}.
\bibitem{PiGPD12} J. M. M. Chavez et al.,
\href{https://doi.org/10.1103/PhysRevD.105.094012} {Phys. Rev. D {\bf 105}, 094012 (2022)}.
\bibitem{PiGPD13} K. Raya et al.,
\href{https://doi.org/10.1088/1674-1137/ac3071} {Chin. Phys. C {\bf 46}, 013105 (2022)}.
\bibitem{PiGPD14} Y. Guo, X. Ji and K. Shiells,
\href{https://doi.org/10.1007/JHEP09(2022)215} {JHEP {\bf 09}, 215 (2022)}.
\bibitem{LFWFs1} G. R. Goldstein, J. O. Hernandez and S. Liuti,
\href{https://doi.org/10.1103/PhysRevD.84.034007} {Phys. Rev. D {\bf 84}, 034007 (2011)}.
\bibitem{LFWFs2} J. O. Hernandez, S. Liuti, G. R. Goldstein and K. Kthuria,
\href{https://doi.org/10.1103/PhysRevC.88.065206} {Phys. Rev. C {\bf 88}, 065206 (2013)}.
\bibitem{LFWFs3} N. Kumar, C. Mondal and N. Sharma,
\href{https://doi.org/10.1140/epja/i2017-12433-0} {Eur. Phys. J. A {\bf 53}, 237 (2017)}.
\bibitem{LFWFs4} B. Kriesten,
\href{https://doi.org/10.1103/PhysRevD.105.056022} {Phys. Rev. D {\bf 105}, 056022 (2022)}.
\bibitem{LFWFs5} Y. Liu et al. (BLFQ Collaboration),
\href{https://doi.org/10.1103/PhysRevD.105.094018} {Phys. Rev. D {\bf 105}, 094018 (2022)}.
\bibitem{Bakulev2000} A. P. Bakulev, R. Ruskov, K. Goeke and N. G. Stefanis,
\href{https://doi.org/10.1103/PhysRevD.62.054018} {Phys. Rev. D {\bf 62}, 054018 (2000)}.
\bibitem{Goeke2001} K. Goeke, M. V. Polyakov and M. Vanderhaeghen,
\href{https://doi.org/10.1016/S0146-6410(01)00158-2} {Prog. Part. Nucl. Phys. {\bf 47}, 401 (2001)}.
\bibitem{Diehl2003} M. Diehl,
\href{https://doi.org/10.1016/j.physrep.2003.08.002} {Phys. Rep. {\bf 388}, 41 (2003)}.
\bibitem{Belitsky2005} A. V. Belitsky and  A. V. Radyushkin,
\href{https://doi.org/10.1016/j.physrep.2005.06.002} {Phys. Rep. {\bf 418}, 1 (2005)}.
\bibitem{Boffi2007} S. Boffi and  B. Pasquini,
\href{https://doi.org/10.1393/ncr/i2007-10025-7} {Riv. Nuovo Cim. {\bf 30}, 387 (2007)}.
\bibitem{PiChi1} W. Broniowski and E. Ruiz Arriola,
\href{https://doi.org/10.1103/PhysRevD.78.094011} {Phys. Rev. D {\bf 78}, 094011 (2008)}.
\bibitem{PiChi2} H. D. Son and H. C. Kim,
\href{https://doi.org/10.1103/PhysRevD.90.111901} {Phys. Rev. D {\bf 90}, 111901 (2014)}.
\bibitem{PiFF1} W. Broniowski, E. Ruiz Arriola and P. Sanchez-Puertas,
\href{https://doi.org/10.1103/PhysRevD.106.036001} {Phys. Rev. D {\bf 106}, 036001 (2022)}.
\bibitem{PiFF2} A. F. Krutov and V. E. Troitsky,
\href{https://doi.org/10.1103/PhysRevD.103.014029} {Phys. Rev. D {\bf 103}, 014029 (2021)}.
\bibitem{PiFF3} Z. Xing, M. Ding and L. Chang,
\href{https://doi.org/10.1103/PhysRevD.107.L031502} {Phys. Rev. D {\bf 107}, L031502 (2023)}.
\bibitem{PiFF4} Y. Z. Xu et al.,
\href{https://doi.org/10.1140/epjc/s10052-024-12518-x} {Eur. Phys. J. C {\bf 84}, 191 (2024)}.
\bibitem{Raya2024} I. M. Higuera-Angulo, R. J. Hernandez-pinto, K. Raya and A. Bashir,
\href{https://doi.org/10.1103/PhysRevD.110.034013}{Phys. Rev. D {\bf 110} 034013 (2024)}.
\bibitem{GA22} H. Hashamipour, M. Goharipour, K. Azizi and V. Goloskokov,
\href{https://doi.org/10.1103/PhysRevD.105.054002} {Phys. Rev. D {\bf 105}, 054002 (2022)}.
\bibitem{GA231} H. Hashamipour, M. Goharipour, K. Azizi and V. Goloskokov,
\href{https://doi.org/10.1103/PhysRevD.107.096005} {Phys. Rev. D {\bf 107}, 096005 (2023)}.
\bibitem{GA232} F. Irani, M. Goharipour, H. Hashamipour and K. Azizi,
\href{https://doi.org/10.1103/PhysRevD.108.074018} {Phys. Rev. D {\bf 108}, 074018 (2023)}.
\bibitem{GA24} M. Goharipour, H. Hashamipour, F. Irani and K. Azizi,
\href{https://doi.org/10.1103/PhysRevD.109.074042} {Phys. Rev. D {\bf 109}, 074042 (2024)}. 
\bibitem{NYVPi} H. Nematollahi and M. M. Yazdanpanah,
\href{https://doi.org/10.1016/j.nuclphysa.2018.05.009} {Nucl. Phys. A {\bf 977}, 23 (2018)}.
\bibitem{Chi1} K. Suzuki and W. Weise,
\href{https://doi.org/10.1016/S0375-9474(98)00132-8} {Nucl. Phys. A {\bf 634}, 141 (1998)}.
\bibitem{Chi2} Y. Ding, R-G. Xu and B-Q. Ma,
\href{https://doi.org/10.1103/PhysRevD.71.094014} {Phys. Rev. D {\bf 71}, 094014 (2005)}.
\bibitem{Chi3} H. Song, X. Zhang and B-Q. Ma,
\href{https://doi.org/10.1140/epjc/s10052-011-1542-4} {Eur. Phys. J. C {\bf 71}, 1542 (2011)}. 
\bibitem{Chi4} H. Nematollahi, M. M. Yazdanpanah and A. Mirjalili,
\href{https://doi.org/10.1088/0954-3899/39/4/045009} {J. Phys. G: Nucl. Part. Phys. {\bf 39}, 045009 (2012)}.
\bibitem{Chi5} A. Watanabe, C. W. Kao and K. Suzuki,
\href{https://doi.org/10.1103/PhysRevD.94.114008} {Phys. Rev. D {\bf 94}, 114008 (2016)}.
\bibitem{Chi6} A. Watanabe, T. Sawada and C. W. Kao,
\href{https://doi.org/10.1103/PhysRevD.97.074015} {Phys. Rev. D {\bf 97}, 074015 (2018)}.
\bibitem{NYSGPi} H. Nematollahi and M. M. Yazdanpanah,
\href{https://doi.org/0.1140/epjp/i2019-12844-2} {Eur. Phys. J. Plus {\bf 134}, 382 (2019)}.
\bibitem{CMMR} N. Chouika, C. Mezrag, H. Moutarde and J. Rodríguez-Quintero,
\href{https://doi.org/10.1016/j.physletb.2018.02.070} {Phys. Lett. B {\bf 780}, 287 (2018)}.
\bibitem{AP} G. Altarelli and G. Parisi,
\href{https://doi.org/10.1016/0550-3213(77)90384-4} {Nucl. Phys. B {\bf 126}, 298 (1977)}.
\bibitem{JLAB} G. M. Huber et al.,
\href{https://doi.org/10.1103/PhysRevC.78.045203} {Phys. Rev. C {\bf 78}, 045203 (2008)}.
\bibitem{PiRad1} P. Zyla et al.,
\href{https://doi.org/10.1093/ptep/ptaa104} {PTEP {\bf 2020}, 083C01 (2020)}.
\bibitem{PiRad2} Z. F. Cui et al.,
\href{https://doi.org/10.1016/j.physletb.2021.136631} {Phys. Lett. B {\bf 822}, 136631 (2021)}.

\end{thebibliography}
\end{document}